\documentstyle[aps]{revtex}
\begin{document}  

\twocolumn[\hsize\textwidth\columnwidth\hsize\csname 
@twocolumnfalse\endcsname

\title{General Relativity in terms of Dirac Eigenvalues} 
\author{Giovanni Landi${}^1$, Carlo Rovelli${}^2$} 
\address{\it 
${}^1$ Dipartimento di Scienze Matematiche, Universit\`a di Trieste, 
I-34127, Trieste, Europe \\ 
${}^1$ INFN, Sezione di Napoli, I-80125 Napoli, Europe \\ 
${}^2$ Physics Department, University of Pittsburgh, 
Pittsburgh 
Pa 15260, USA \\ 
${}^2$ Center for Gravity and Geometry, Penn State 
University, State College Pa 16801, USA
}
\maketitle

\begin{abstract}
 
The eigenvalues of the Dirac operator on a curved spacetime are 
 diffeomorphism-invariant functions of the geometry.  They 
form an infinite set of ``observables'' for general relativity.  
Recent work of Chamseddine and Connes suggests that they can be taken 
as variables for an invariant description of the gravitational 
field's dynamics.  We compute the Poisson brackets of these eigenvalues 
and find them in terms of the energy-momentum of the eigenspinors and the 
propagator of the linearized Einstein equations.  We show that the 
eigenspinors' energy-momentum is the Jacobian matrix of the change of 
coordinates from metric to eigenvalues.  We also consider a minor 
modification of the spectral action, which eliminates the disturbing 
huge cosmological term and derive its equations of motion. These are  
satisfied if the energy momentum of the trans Planckian 
eigenspinors scale linearly with the eigenvalue; we argue that this 
requirement approximates the Einstein equations. 
\vskip.5cm
{gr-qc/yymmnn \hskip7cm\today} 
\end{abstract}
\vskip1cm 
]

\section{Introduction}

One of the important lessons that we learn from general relativity is that 
fundamental physics is invariant under diffeomorphisms: there is no 
fixed nondynamical structure with respect to which location or motion  
could be defined.  A fully diffeomorphism-invariant description of the 
geometry has consequently long been searched in general relativity; 
but so far without much success.  As emphasized by many authors, such 
a description would be precious for quantum gravity \cite{diff}.

Recently, Alain Connes' intriguing attempt of using the particle physics
standard model for unraveling a microscopic noncommutative structure of
spacetime \cite{alainb,daniel3,landi}, has generated --in a sense as a 
side product-- the remarkable idea that the curved-spacetime Dirac 
operator $D$ codes the full information about spacetime geometry 
in a way that can be used for describing the dynamics of the latter.  
Indeed, not only the geometry can be reconstructed from the (normed)
algebra  generated by (the inverse of) $D$ and the smooth functions on
the manifold, but the  Einstein-Hilbert action itself is approximated  by
the trace of  a simple function of $D$ \cite{alain}. In its simplest reading,
this  result suggests the possibility of taking the eigenvalues 
$\lambda^{n}$ of the Dirac operator as ``dynamical variables'' for 
general relativity.  Indeed, these form an infinite family of fully 
four-dimensional diffeomorphism invariant observables, precisely the 
kind of object that was long searched in relativity.  This approach has 
limitations.  The most serious of these are that so far it works for 
Euclidean general relativity only (see \cite{eli} for some attempts to 
overcome this problem), and somewhat unrealistic predictions for the 
bare couplings \cite{daniel2}.  However, it definitely opens 
a new window on the study of the dynamics of spacetime.

In order to use these ideas in the classical or in the quantum theory, 
one must translate structures from the metric variables to the 
$\lambda^{n}$ variables.  In particular, one needs information on the 
constraints that the $\lambda^{n}$'s satisfy if they correspond to a 
smooth geometry, and on their Poisson brackets.  The difficulty is 
that the dependence of the $\lambda^{n}$'s on the geometry is defined 
in a very implicit manner, and these quantities seem too hard to 
compute.

In this letter, we show that these difficulties can be circumvented. 
Following some earlier results in \cite{roberto} (valid only for the non
4d-invariant eigenvalues of the fixed-time Weyl operator), 
we derive here an expression for the
Poisson brackets of the Dirac eigenvalues.  Rather surprisingly, this
expression turns out to be given in terms of the energy-momentum tensors
of the Dirac eigenspinors.  These tensors form the matrix elements of the
Jacobian matrix of the change of coordinates between metric and
eigenvalues. The Poisson brackets are quadratic in these tensors, with a
kernel given by the propagator of the linearized Einstein equations.  
Thus, the energy-momentum tensors of the Dirac eigenspinors turn out 
to be the key tool for analyzing the representation of spacetime 
geometry in terms of Dirac eigenvalues.

We study also the Chamseddine-Connes spectral action.  In the form 
presented in \cite{alain} this is quite unrealistic as a pure gravity 
action, because of a huge cosmological constant term that implies that 
the geometries for which the action approximates the Einstein-Hilbert 
action are {\em not} solutions of the theory.  However, a very small 
modification of the action eliminates the cosmological constant term.  
We derive the equations of motion directly from the (modified) spectral 
action.  We argue that they amount to the requirement that the energy 
momenta of the high mass eigenspinors scale linearly with the mass, 
and that this requirement approximates the vacuum Einstein equations.

These results suggest that --even independently from its application 
to the standard model-- the Chamseddine-Connes {\em gravitational} 
theory can be viewed as a manageable gravitational theory by itself 
(see also \cite{fr,gianni}), possibly with powerful applications to 
classical and quantum gravity.  It reproduces general relativity at 
low energies; it is formulated in terms of fully diffeomorphism 
invariant variables; and, of course, it prompts fascinating extensions of 
the very notion of geometry.

\section{GR in terms of eigenvalues}

Consider Euclidean general relativity (GR) on a compact 4d (spin-) manifold
without boundary.   We work in terms of the tetrad field $E_{\mu}^{I}(x)$.  Here 
$\mu=1\ldots 4$ are spacetime indices and $I=1\ldots 4$ are internal 
Euclidean indices, raised and lowered by the Euclidean metric 
$\delta_{IJ}$.  The metric field is given by $g_{\mu\nu}(x)= 
E_{\mu}^{I}(x)E_{\nu\, I}(x)$, and is used to raise and lower 
spacetime indices.  The spin connection $\omega_{\mu}^{I}{}_{J}$ is 
defined by the equation $\partial_{[\mu}E_{\nu]}^{I} = 
\omega_{[\mu}^{I}{}_{J} E_{\nu]}^{J}$.  The dynamics is defined by the 
action $S[E]=\frac{1}{16\pi G} \int d^{4} x \sqrt{g} R$, where $g$ and 
$R$ are the determinant and the Ricci scalar of the metric.

In spite of a curiously widespread popular belief of the contrary, 
phase space is a covariant notion: the covariant definition of the 
phase space is as the space of the solutions of the equations of 
motion, modulo gauge transformations \cite{phase}.  In the theory 
considered, the gauge transformations are given by 4d diffeomorphisms 
and by the local internal rotations of the tetrad field.  Thus, the 
phase space $\Gamma$ of GR is the space of the tetrad fields $E$ that 
solve the equation of motion (Einstein equations), modulo internal 
rotations and diffeomorphisms.  $\Gamma$ can be identified with the space of the 
gauge orbits on the constraint surface and with the space of the Ricci 
flat 4-geometries.

We shall use the following notation.  We denote the space of smooth 
tetrad fields as $\cal E$; the space of the orbits of the gauge groups 
--diffeomorphisms and local orthogonal tetrad rotations-- in $\cal E$ 
as $\cal G$ (these are ``4-geometries''); and the space of the orbits 
that satisfy the Einstein equation as $\Gamma$ (these are the Ricci-flat 
4-geometries, which form the phase space of GR).  We call functions 
on $\Gamma$ ``observables''.  Observables correspond to functions on 
the constraint surface that commute with {\em all\/} the constraints.

We now define an infinite family of such observables. Consider spinor 
fields $\psi$ on the manifold and the curved Dirac operator
\begin{equation} 
D = \imath \gamma^{I} E^{\mu}_{I} \left(\partial_{\mu}+\omega_{\mu\, 
JK}\gamma^{J} \gamma^{K}\right),
\end{equation}
which acts on them. Here $\gamma$ are the (Euclidean) hermitian Dirac 
matrices.  For each given field $E$, the Dirac operator is 
a self-adjoint operator on the Hilbert space of spinor fields with 
scalar  product
\begin{equation}
(\psi,\phi)=\int d^{4}x \sqrt{g}\ \overline{\psi(x)} \phi(x).
\label{product}
\end{equation}
where the bar indicates complex conjugation, and the scalar 
product in spinor space is the natural one in $C^{4}$.  Therefore, $D$ 
admits a complete set of real eigenvalues and eigenfunctions 
(``eigenspinors'').  Since the manifold is compact, the spectrum is 
discrete.  We write
\begin{equation}
D \psi^{n} = \lambda^{n} \psi^{n}.
\end{equation}
Here and below, $n=0,1,2 ... $ is an index, not an exponent.  We 
convene to label the eigenvalues so that $\lambda^{n}$ is non 
decreasing in $n$, namely $\lambda^n\leq\lambda^{n+1}$ (each eigenvalue is
repeated according to its multiplicity).  In order to  emphasize the dependence
of Dirac operator, eigenvalues and  eigenspinors on the tetrad field, we use
also the notation
\begin{equation}
D[E]\  \psi^{n}[E] = \lambda^{n}[E]\  \psi^{n}[E]
\end{equation}
where the dependence on the tetrad is indicated explicitly.

$\lambda^{n}[E]$ defines a discrete family of real 
functions on the space $\cal E$ of the tetrad fields
 \begin{eqnarray}
 \lambda^{n}:\ \   {\cal E} & \longrightarrow & R        \nonumber \\
		E & \longmapsto & \lambda^{n}[E]. 
 \end{eqnarray}
Equivalently, they define a function $\lambda$ from $\cal E$ into 
the space of infinite sequences $R^\infty$
 \begin{eqnarray}
 \lambda:\ \   {\cal E} & \longrightarrow & R^\infty      \nonumber \\
		E & \longmapsto & \{\lambda^{n}[E]\}.
 \end{eqnarray}
Since we have chosen to order the $\lambda^n$'s in non-decreasing 
order, the image of $\cal E$ under this map, which we denote as 
$\lambda({\cal E})$ is entirely contained in the cone $\lambda^n \leq 
\lambda^{n+1}$ of $R^\infty$. 

Now, for every $n$, the function $\lambda^{n}$ is invariant under 4d 
diffeomorphisms and under internal rotations of the tetrad field.  
Therefore the functions $\lambda^{n}$ are well defined functions on 
$\cal G$.  In particular, they are well defined on $\Gamma$: they are 
observables of GR.  

Two metric fields with the same $\lambda^{n}$'s are called
``isospectral''.  Isometric (that is, gauge equivalent) $E$ fields are
isospectral, but the converse might fail to be true \cite{drum,roberto}. 
Therefore $\lambda$ may not be injective even if restricted to $\cal G$. 
The $\lambda^n$'s may fail to coordinatize $\cal G$.  They may also
fail to coordinatize $\Gamma$.  However, they presumably ``almost do it''. 
In the following, we explore the idea of analyzing GR in terms of the set
of observables $\lambda^{n}[E]$. 

\section{Symplectic structure}

The phase space $\Gamma$ is a symplectic manifold and a Poisson brackets 
structure is defined on the physical observables.  We now study the 
Poisson brackets $\{\lambda^{n}, \lambda^{m}\}$.

To this purpose, we first construct the symplectic structure on $\Gamma$.  This 
can be written in covariant form following Ref~\cite{abhay}.  A vector 
field $X$ on $\Gamma$ can be written as a differential operator
\begin{equation}
X  =  \int d^{4}x \  X_{\mu}^{I}(x)[E]\ \ \frac{\delta}{\delta 
E_{\mu}^{I}(x)} 
\end{equation}
where $X_{\mu}^{I}(x)[E]$ is any solution of the Einstein equations 
{\it linearized\/} over the background $E$.  The symplectic two-form 
$\Omega$ of GR is given by
\begin{equation}
\Omega(X,Y) = \frac{1}{32\pi G} \int_{\Sigma}d^{3}\sigma\  n_{\rho}
[ 
X_{\mu}^{I}\ \overleftarrow{\overrightarrow{\nabla}}{}_{\tau}\ Y_{\nu}^{J} 
] \epsilon^{\tau}_{IJ\upsilon} \epsilon^{\upsilon\rho\mu\nu} 
\label{ome}
\end{equation}
where $[X_{\mu}^{I}\ \overleftarrow{\overrightarrow{\nabla}}{}_{\tau}\ 
Y_{\nu}^{J}] \equiv [X_{\mu}^{I}\ \nabla_{\tau}\ Y_{\nu}^{J} - 
Y_{\mu}^{I}\ \nabla_{\tau}\ X_{\nu}^{J}]$; from now on we put $32\pi G=1$. 
Both sides of (\ref{ome}) are 
functions of $E$, namely scalar functions on $\Gamma$; this $E$ is 
used to transform internal indices into spacetime indices.  Here 
$\Sigma:\sigma\longmapsto x(\sigma)$ is an arbitrary three-dimensional 
``ADM'' surface, and $n_{\rho}$ is the normal one-form to this surface.  
The coefficients of the symplectic form can be written as
\begin{equation}
\Omega^{\mu\nu}_{IJ}(x,y) = \! \int_{\Sigma}\! d^{3}\sigma\ n_{\rho} 
[\delta(x,x(\sigma)) \overleftarrow{\overrightarrow{\nabla}}_{\tau}  
\delta(y, x(\sigma)) ] \epsilon^{\tau}_{IJ\upsilon} 
\epsilon^{\upsilon\rho\mu\nu} .
\label{omega}
\end{equation}

The Poisson bracket between two functions $f$ and $g$ on $\Gamma$ is  
given by
\begin{equation}
\{f,g\}= \int d^{4}x\int d^{4}y\ \ P_{\mu\nu}^{IJ}(x,y) \ 
\frac{\delta f}{\delta E_{\mu}^{I}(x)}\ 
\frac{\delta g}{\delta E_{\mu}^{I}(y)}.
\label{pp}
\end{equation}
where $P_{\mu\nu}^{IJ}(x,y)$ is the inverse of the symplectic form 
matrix.  It is defined by
\begin{equation}
\int d^{4}y\ P_{\mu\nu}^{IJ}(x,y)\ \Omega^{\nu\rho}_{JK}(y,z)=
\delta(x,z)\ \delta_{\mu}^{\rho}\ \delta^{I}_{K}.
\end{equation}
Since the symplectic form is degenerate on the space of the fields 
(it is non-degenerate only when restricted to the space 
of equivalent classes of gauge-equivalent fields), we can only invert 
it on this space by fixing a gauge.  Let us assume this has been done.  
More precisely, integrating the last equation against a vector field 
$F_{\rho}^{K}(z)$ that satisfies the linearized Einstein equations 
over $E$, we have
\begin{eqnarray}
&\int d^{4}y \int d^{4}z\ P_{\mu\nu}^{IJ}(x,y)\ 
\Omega^{\nu\rho}_{JK}(y,z) \ 
F_{\rho}^{K}(z) = & \nonumber \\
& \ \ \ = \int d^{4}z \ \delta(x,z)\ 
\delta_{\mu}^{\rho}\ \delta^{I}_{K} \ F_{\rho}^{K}(z),&
\end{eqnarray}
Integrating over the delta functions, and using (\ref{omega}), we have
\begin{eqnarray}
&\int_{\Sigma} d^{3}\sigma \ n_{\rho} [ 
P^{IJ}_{\mu\nu}(x,x(\sigma)) \overleftarrow{\overrightarrow{\nabla}}_{\rho} 
F^{K}_{\tau}(x(\sigma)) ]
\epsilon^{\rho}_{JK\upsilon}\epsilon^{\upsilon\nu\tau\sigma} = &
\nonumber \\
& \ \ \ \ \ = F_{\mu}^{I}(x).&
\label{p}
\end{eqnarray}
This equation, where $F$ is any solution of the linearized equations, 
defines $P$.  But this equation is precisely the definition of the 
propagator of the linearized Einstein equations over the background 
$E$ (in the chosen gauge).  For instance, let us chose the surface 
$\Sigma$ as $x^{4}=0$ and fix the gauge with
\begin{equation}
X^{4}_{4}  =  1, \ \ 
X^{4}_{a} = 0, \ \ 
 X^{i}_{4} = 1, \ \  
 X^{i}_{a} = 0.
\end{equation}
where $a=1,2,3$ and $i=1,2,3$.  Then equation (\ref{p}) becomes
\begin{equation}
F^{i}_{a}(\vec x, t)=
\int d^3\vec y\ (P^{ib}_{aj}(\vec x, t; \vec y, 0) 
\overleftarrow{\overrightarrow{\nabla}}_{0} F^{j}_{b}(\vec y, 0)),
\end{equation}
where we have used the notation $\vec x = (x^1, x^2, x^3)$ and 
$t=x^4$, and the propagator can be easily recognized. 

Next, we need the functional derivative of the eigenvalues
with respect to the metric. 
The  variation of $\lambda^{n}$ for a variation of $E$ can be computed 
using standard techniques for first order variations of eigenvalues;  
(standard time-independent quantum-mechanics perturbation theory).  For 
a self-adjoint operator $D$ depending on a parameter $v$ and whose 
eigenvalues are nondegenerate, we have
\begin{equation}
\frac {d\lambda^{n}}{d v} = (\psi ^{n}| \frac{d}{dv } D(v)|\psi^{n}).
\end{equation}
In our situation, for generic metrics with nondegenerate eigenvalues 
we have that
\begin{eqnarray}
\frac{\delta \lambda^{n}}{\delta E_{\mu}^{I}(x)} & = & 
(\psi^{n}|\frac{\delta}{\delta E_{\mu}^{I}(x)}D|\psi^{n}) \\
	 & = & 
\int \sqrt{g}\ \bar\psi^{n}  \frac{\delta}{\delta E_{\mu}^{I}(x)} 
D\psi^{n} 
\nonumber 
\\ & = & 
\frac{\delta}{\delta E_{\mu}^{I}(x)} \int \sqrt{g}\ \bar\psi^{n}  
D\psi^{n} - \int \frac{\delta\sqrt{g}}{\delta E_{\mu}^{I}(x)} 
\bar\psi^{n}  D \psi^{n}
\nonumber 
\\ & = &
\frac{\delta}{\delta E_{\mu}^{I}(x)} \int \sqrt{g}\ \bar\psi^{n}  
D\psi^{n} - \int \frac{\delta\sqrt{g}}{\delta E_{\mu}^{I}(x)} 
\bar\psi^{n} \lambda^n \psi^{n}
\nonumber 
\\ & = &
\frac{\delta}{\delta E_{\mu}^{I}(x)} \int \sqrt{g}\ 
(\bar\psi^{n}  D\psi^{n} - \lambda^n \bar\psi^{n}\psi^n) 
\nonumber 
\\
& = & T^{n}{}^{\mu}_{I}(x).
\label{t}
\end{eqnarray}
Now, $T^{n}{}^{\mu}_{I}(x)$ is nothing but the usual energy-momentum 
tensor of the spinor field $\psi^{n}$ in tetrad notation (see for 
instance \cite{stanley}).  Indeed, the usual Dirac energy-momentum 
tensor is defined by
\begin{equation}
T^{\mu}_{I}(x)\equiv\frac{\delta}{\delta E_{\mu}^{I}(x)}S_{\rm Dirac},
\end{equation}
where $S_{\rm Dirac}=\int \sqrt{g}\ (\bar 
\psi D \psi - \lambda\bar\psi\psi)$ is the usual  
Dirac action of a spinor with mass $\lambda$.

We have shown that the energy-momentum tensor of the 
eigenspinors 
gives the Jacobian matrix of the transformation from $E$ to $\lambda$; 
namely it gives the variation of the eigenvalues for a small 
change in the geometry. This fact suggests that we can study the map $\lambda$ 
locally by studying the space of the eigenspinor's energy-momenta.

By combining (\ref{pp},\ref{p}) and (\ref{t}) we obtain our main result: 
\begin{equation}
\{\lambda^{n},\lambda^{m}\}= 
\int\!\! d^{4}x\!\!  
\int\!\! d^{4}y \ T^{[n}{}^{\mu}_{I}(x)\ P_{\mu \nu}^{IJ}(x, y) \ 
T^{m]}{}^{\nu}_{J}(y)
\label{main}
\end{equation}
which gives the Poisson bracket of two eigenvalues of the Dirac operator 
in terms of the energy-momentum tensor of the two corresponding 
eigenspinors and of the propagator of the linearized Einstein 
equations.  The r.h.s.  does not depend on the gauge chosen for $P$.

Finally, if the transformation between the ``coordinates'' 
$E_{\mu}^{I}(x)$ and the ``coordinates'' $\lambda^{n}$ is locally 
invertible on the phase space $\Gamma$, we can write the symplectic 
form directly in terms of the $\lambda^{n}$'s as
\begin{equation}
\Omega=\Omega_{mn}\ d\lambda^{n} \wedge d\lambda^{m},
\end{equation}
where a sum over indices is understood, and 
where $\Omega_{mn}$ is defined by
\begin{equation}
\Omega_{mn}\ T^{n}{}^{\mu}_{I}(x)\ T^{m}{}^{\nu}_{J}(y)= 
\Omega^{\mu\nu}_{IJ}(x,y).
\end{equation}
Indeed, let $d E_{\mu}^{I}(x)$ be a (basis) one-form on $\Gamma$, 
namely the infinitesimal difference between two solutions of Einstein 
equations, namely a solution of the Einstein equations linearized 
over $E$.  We have then
\begin{eqnarray} 
\Omega & = & \int d^{4}x \int d^{4}y \ 
\Omega^{\mu\nu}_{IJ}(x,y)\ d E_{\mu}^{I}(x) \wedge d E_{\nu}^{J}(y)
\nonumber \\
	 & = &  \int d^{4}x \int d^{4}y 
\Omega_{mn}\ T^{n}{}^{\mu}_{I}(x)\ d E_{\mu}^{I}(x) \wedge 
T^{m}{}^{\nu}_{J}(y)\ d E_{\nu}^{J}(y) 
\nonumber \\
	 & = & \Omega_{mn} \ 
d\lambda^{n} \wedge d\lambda^{m}.
\end{eqnarray}
An explicit evaluation of the matrix $\Omega_{nm}$ would be of great interest.

\section{Action}

As shown in \cite{alain}, the gravitational action can be expressed as
\begin{equation}
 S = Tr[\chi(D)]
\label{action1}
\end{equation}
in natural units $\hbar=G=c=1$.  Here $\chi$ is a smooth monotonic 
function on $R^{+}$ such that
\begin{eqnarray} 
\chi(x) &=& 1,\ \ 
{\rm for}\ x < 1-\delta, \nonumber \\
\chi(x) &=& 0,\ \ 
{\rm for}\ x> 1 + \delta.
\end{eqnarray}
where $\delta<<<1$. Equivalently, $S$ is the number of $\lambda^n$'s 
smaller than the Planck mass, which is 1 in natural units.

The action (\ref{action1}) approximates the Einstein-Hilbert action 
with a large cosmological term for ``large-scale'' metrics with small 
curvature (with respect to the Planck scale).  This can be seen as follows.  
Let $N[E]$ be the integer such that $\lambda^{N}[E]$ is the largest 
eigenvalue of $D[E]$ smaller than the Planck mass $M_{P}={1\over 
L_{P}}=1$.  A moment of reflection shows then that we have
 \begin{equation}
	S[E]=N[E]
\label{sn}
\end{equation}
by definition. For large $n$, the growth of the eigenvalues of the Dirac
operator is given by the Weyl formula; 
\begin{equation}
  \lambda^n = V^{-{1\over 4}}\ n^{{1\over 4}} + \ldots
  \label{qua}
\end{equation}
where $V$ is the volume and we are neglecting factors of the 
order of unity.  The next term in this asymptotic 
approximation can be obtained from the fact that the Dixmier trace 
(the logarithmic divergence of the trace) of $D^{-2}$ is the 
Einstein-Hilbert action \cite{alainb,alain,landi}; therefore
\begin{equation}
(\lambda_{n})^{-2} = V^{{1\over2}}\ n^{-{1\over 2}} + \int\sqrt{g}R\ 
{n^{-1}} + \ldots 
\end{equation}
Under our assumptions on the geometry, at the Planck scale we are in 
asymptotic regime: the first term dominates over the second, and the 
remaining terms are negligible.  Writing the last equation for $n=N$ 
and using (\ref{sn}), we have
\begin{equation}
1 = V^{{1\over2}}\ S^{-{1\over 2}} + \int\sqrt{g}R\ S^{-1} + \ldots
\end{equation}
Solving for $S$ we obtain 
\begin{equation}
S = \frac{1}{L_P^4} V + \frac{1}{L_P^2} \int\sqrt{g}R + 
\ldots. \end{equation}
which shows that the action (\ref{action1}) is dominated by the 
Einstein-Hilbert action with a cosmological term. In the last 
equation, we have explicitly reinserted the Planck length $L_P$. 

The presence of the huge Planck-mass cosmological term is a bit devastating 
because the solutions of the equations of motions have Planck-scale 
Ricci scalar, and therefore they are {\em all\/} out of the regime for which 
the approximation taken is valid!  

However, the cosmological term can be canceled easily by replacing $\chi(x)$ 
with $\tilde\chi(x)$ defined by
\begin{equation}
\tilde\chi(x)=\chi(x)-\epsilon^{4} \chi(\epsilon x)
\label{epsilon}
\end{equation}
for a small number $\epsilon$.  The additional term cancels exactly 
the cosmological term, leaving only the Einstein Hilbert action, with 
corrections which are small for low curvature geometries, which, {\em 
now}, {\em are} solutions of the theory.  In fact
\begin{eqnarray}
\tilde S &\equiv & Tr(\tilde{\chi}(D)) \nonumber \\
&=& Tr({\chi}(D)) -\epsilon^{4} 
Tr({\chi}(\epsilon D)) \nonumber \\
& =& \frac{V}{L_{P}^4} +  \frac{1}{L_{P}^2} \int\sqrt{g}R \nonumber \\
& & -\epsilon^{4}\left(\epsilon^{-4}  \frac{V}{L_{P}^4} 
+\epsilon^{-2}  \frac{1}{L_{P}^2} \int\sqrt{g}R\right) + \ldots.  
\nonumber \\ &=&  \frac{1}{L_{P}^2}\ \int\sqrt{g} R + \ldots
\end{eqnarray}

If we write $S$ directly in terms of the observables $\lambda^{n}$, we 
have the following expression for the action
\begin{equation}
 \tilde S[\lambda] = \sum_n\ \tilde\chi(\lambda^n).
 \label{action}
\end{equation}
One cannot vary the $\lambda^{n}$'s in this expression to 
obtain (approximate) Einstein equations.  These equations are obtained 
minimizing (\ref{action}) on the surface $\lambda({\cal E})$, not on 
the entire $R^{\infty}$.  In other words, the $\lambda^{n}$'s are not 
independent: there are relations among them.  These relations code the 
complexity of GR.  The equations of motion are obtained by varying $S$ 
with respect to the tetrad field.  They can be written as
\begin{equation}
0  =  \frac{\delta \tilde S}{\delta E_{\mu}^{I}(x)} 
 =  \sum_n\ \frac{\partial \tilde S}{\partial \lambda^{n}} \ 
\frac{\delta \lambda^{n}}{\delta E_{\mu}^{I}(x)}
 =  \sum_n\ \frac{d\tilde\chi(\lambda^n)}{d \lambda^{n}}\ 
T^{n}{}^{\mu}_{I}(x).
\label{ee}
\end{equation}
Let $f(x) = \frac{d}{dx}\tilde\chi(x)$.  The equations of motion of the 
theory are then
\begin{equation}
\sum_{n} f(\lambda^{n})\ T^{n}{}^{\mu}_{I}(x) = 0. 
\end{equation}

We close by analyzing the content of these equations.  $f(x)$ is a 
function that vanishes everywhere except on two narrow peaks.  A 
positive peak (width $\delta$ and height $1/\delta$) around the Planck 
mass 1; and a negative peak (width $\delta\epsilon$ and height 
$\epsilon^{5}/\delta$) around the arbitrary large number 
$s={1\over\epsilon}>>1$.  The equation is therefore solved if above 
the Planck mass ($n>>N$), the energy momentum tensor scales as
\begin{equation}
\rho(1)\ T^{N}{}^{\mu}_{I}(x) = s^{-4}\ \rho(s)\ 
T^{N(s)}{}^{\mu}_{I}(x),
\end{equation}
where $\rho(s)$ is the density of the eigenvalues at the scale $s$ and 
$\lambda^{N(s)}=s$, because in this case the two terms from the two 
peaks cancel.  But from (\ref{qua}) we have that the density of 
eigenvalues grows as $N^{3}$, and that $N(s)=s^{4}$. This yields
\begin{equation}
T^{n}{}^{\mu}_{I}(x) = \lambda^{n}\ \ T^{N}{}^{\mu}_{I}(x).
\label{scaling}
\end{equation}
for $n>>N$.  (Recall that $\lambda^{N}=1$.) 
In other words: {\it the equations of motion for the geometry are solved 
when above the Planck mass the energy-momentum of the eigenspinors 
scales as the mass}.

To understand the meaning of this scaling requirement, notice that 
$T^{n}{}_{\mu}^{I}$ is formed by a term linear in the derivatives of 
the spinor field and a term independent from these, which is a function 
of $(\psi, E, \partial_{\mu} E)$ quadratic in $\psi$.
\begin{equation}
T^{n}{}_{\mu}^{I}=\bar\psi^{n}\gamma^{I}
{\scriptstyle \overleftarrow{\overrightarrow{\partial}}}_{\mu}\psi^{n} +
S^{n}{}_{\mu}^{I}[\psi, E, \partial  E].
\end{equation}
If we expand the last term around a point of the manifold with local coordinates 
$x$, covariance and dimensional  analysis require that
\begin{equation}
S^{n}{}_{\mu}^{I} = 
c_{0} \lambda^{n} E_{\mu}^{I}+c_{1}\ R_{\mu}^{I}+ c_{2}\  
R\,E_{\mu}^{I}+O\left( \frac{1}{\lambda^{n}}\right).
\end{equation}
for some fixed expansion coefficients $c_{0}, c_{1}$ and $c_{2}$.  
Here $R_{\mu}^{I}$ is the Ricci tensor.  [To be convinced that terms 
of this form do appear, consider the following.
\begin{eqnarray}
T^{n}{}_{\mu}^{I} &=& \bar\psi^{n}\gamma^{I}D_{\mu}\psi^{n} +\ldots  
\nonumber \\
& = & (\lambda^{n})^{-1}\ 
 \bar\psi^{n} 
\gamma^{I}\gamma^{\nu}D_{\mu}D_{\nu}\psi^{n}+\ldots 
\nonumber \\
& = &  (\lambda^{n})^{-1}\ 
\bar\psi^{n} \gamma^{I}\gamma^{\nu}[D_{\mu},D_{\nu}]\psi^{n}  +\ldots
\nonumber \\
& = & (\lambda^{n})^{-1}\ 
\bar\psi^{n} \gamma^{I}\gamma^{\nu}R_{\mu\nu}\psi^{n} +\ldots
 \nonumber \\
&=&   (\lambda^{n})^{-1}\ 
\bar\psi^{n} 
\gamma^{I}\gamma^{\nu}R_{\mu\nu}^{JK}\gamma_{J}\gamma_{K}\psi^{n} +\ldots
 \nonumber \\ 
& = &
Tr\ \  \gamma^{I}\gamma^{\nu}R_{\mu\nu}^{JK}\gamma_{J}\gamma_{K} +\ldots
 \nonumber \\ 
& = &
R^{I}_{\mu} + \ldots\ ]
\end{eqnarray}
For sufficiently high $n$, the eigenspinors are locally plane waves in 
local cartesian coordinates: doubling the mass just doubles the frequency. 
If 
\begin{equation}
	\lambda^{m}=t\ \lambda^{n}
\label{lsca}
\end{equation}
Then $\partial_\mu\psi^{m}=t\ \partial_\mu\psi^{n}$.  
It follows that in general the energy momentum scales as
\begin{eqnarray}
T^{m}{}_{\mu}^{I} &=& t 
\left[\bar\psi^{n}\gamma^{I}\partial_{\mu}\psi^{n}- 
\partial_{\mu}\bar\psi^{n}\gamma^{I}\psi^{n} + c_{0} 
\lambda^{n}E_{\mu}^{I}\right] \nonumber \\
& &  + \left[c_{1}\ R_{\mu}^{I}+ c_{2}\ R\,E_{\mu}^{I}\right] + 
O\left(\frac{1}{\lambda^{n}}\right).
\end{eqnarray}
For large $\lambda^{n}$ we can disregard the last term, and (\ref{scaling})  
requires that the second square bracket vanishes.  Taking the trace we 
have $R=0$, using which we conclude
\begin{equation}
                 R^{I}_{\mu} = 0
\end{equation}
which are the vacuum Einstein equations.  Thus, the variation of the 
(modified) Chamseddine-Connes action implies a scaling requirement on the 
high mass eigenspinors energy momentum tensors, and this requirement, 
in turn, yields vacuum Einstein equations at low scale.

\vskip 1cm

We thank Alain Connes, Roberto De Pietri, J\"urg Fr\"ohlich and Daniel 
Kastler for suggestions and conversations.  This work was supported by 
the Italian MURST and by NSF grant PHY-5-3840400.


\begin{thebibliography}{99}

\bibitem{diff} CJ Isham, ``Structural Issues in Quantum Gravity'', 
Imperial/TP/95-96/07, gr-qc/9510063; to appear in the proceedings of the GR14.

\bibitem{alainb} A Connes, {\em Non commutative geometry}, Academic Press
1994; J Math Phys 36 (1995) 6194; A Connes, J Lott, Nucl Phys Supp B18
(1990) 295; in {\em Proceedings of the 1991 Carg\'ese summer school},
edited by J Fr\"ohlich et al, Plenum, New York 1992. 

\bibitem{daniel3} D Kastler, Rev Math Phys 5 (1993) 477; Rev Math Phys 8 
(1996) 103; D Kastler, T Sch\"ucker, ``A detailed account of Alain 
Connes`s version of the standard model in non-commutative differential 
geometry IV'', hep-th/9501177, in print on Rev Math Phys; ``The 
Standard Models \`a la Connes-Lott'', hep-th/9412185; B Iochum, D 
Kastler, T Sch\"ucker, ``Fuzzy mass relations for the Higgs'', 
hep-th/9506044, to appear in J Math Phys; ``Fuzzy mass relations in 
the standard model'', hep-th/9507150; ``Riemannian and Non-commutative 
Geometry in Physics'', hep-th/9511011.

\bibitem{landi} G Landi, {\it An Introduction to Noncommutative Spaces
and their Geometry}, lectures notes, to appear.  

\bibitem{alain} A Connes ``Gravity coupled with matter and the foundation
of non commutative geometry'', hep-th/9603053, AH Chamseddine, A Connes,
``The Spectral Action Principle'', hep-th/9606001; ``A Universal Action 
formula'', hep-th/9606056. 

\bibitem{eli} M Kalau, J Geom Phys 18 (1996) 349;
E Hawkins, ``Hamiltonian Gravity and Noncommutative 
Geometry'', Penn State preprint CGPG-96/5-8, gr-qc/9605068.

\bibitem{daniel2} B Iochum, D Kastler, T Sch\"ucker, ``On the Universal 
Chamseddine-Connes action'', hep-th/9607158. 

\bibitem{roberto} R De Pietri, C Rovelli, Classical and Quantum Gravity 
12, 1279 (1995). 

\bibitem{fr} D Kastler, Comm Math Phys 166 (1995) 633; W Kalau, M 
Walze, J of Geom and Phys 16 (1995) 327; AH Chamseddine, G Felder, J 
Fr\"ohlich, Comm Math Phys 155 (1993) 109, 205; AH Chamseddine, J 
Fr\"ohlich, O Grandjean, J Math Phys 36 (1995) 6255;
M Seriu, Phys Rev D53 (1996) 6902.

\bibitem{gianni} G Landi, C Rovelli, ``Kinematic and Dynamics of 
General Relativity with Noncommutative Geometry'', in preparation.

\bibitem{phase} JL Lagrange, Mem Cl Sci Math Phys Inst France (1808) pg 1; 
JM Souriau, {\em Structure des syst\`emes dynamiques}, Dunod Paris 1970, 
pp. 132, 146; P Chernoff, J Mardsen, {\em Infinite Dimensional Hamiltonian 
Systems}, Springer Lecture Notes in Mathematics, Vol 452, Berlin 1974; BF 
Schutz, {\em Geometrical Methods in Mathematical Physics}, Cambridge 
University Press, Cambridge 1980; $\check{C}$ Crnkovi\'c, E Witten, in {\em 
Newton's tercentenary volume}, edited by SW Hawking and W Israel, 
Cambridge University Press 1987. 

\bibitem{drum} PB Gilkey, {\it Invariance theory, the heat equation and
the Atiyah-Singer index theorem}, p. 330 CRC Press 1995; C Gordon, DL
Webb, S Wolpert, Bull Amer Math Soc. 27 (1992) 134; 
M Kac Ann math Monthly 73 (1966) 1. 

\bibitem{abhay} A Ashtekar, L Bombelli, O Reula, in {\it Mechanics, 
Analysis and Geometry: 200 Years after Lagrange}, edited by M 
Francaviglia, Elsvier 1991.

\bibitem{stanley} S Deser, P van Nieuwenhuizen, Phys Rev D 10 (1974) 411. 

\end{thebibliography}
\end{document}